\documentclass[12pt,preprint]{aastex}

\def\earth{\oplus}
\usepackage{epsfig}
\usepackage{graphicx}
\usepackage{subfigure}
\begin{document}
\title{Stealing the Gas: Giant Impacts and the Large Diversity in Exoplanet Densities}
 \author{Niraj K. Inamdar\altaffilmark{1}  \& Hilke E. Schlichting\altaffilmark{1,2,3} }
 \altaffiltext{1} {Massachusetts Institute of Technology, 77 Massachusetts Avenue, Cambridge, MA 02139, USA} \email{inamdar@mit.edu, hilke@mit.edu}
 \altaffiltext{2} {UCLA, 595 Charles E. Young Drive East, Los Angeles, CA 90095, USA}
 \altaffiltext{3} {California Institute of Technology, 1200 E. California Blvd., Pasadena, CA 91125, USA}

\begin{abstract}
Although current sensitivity limits are such that true Solar System analogs remain challenging to detect, numerous planetary systems have been discovered that are very different from our own Solar System. The majority of systems harbor a new class of planets, bodies that are typically several times more massive than the Earth but that orbit their host stars well inside the orbit of Mercury. These planets frequently show evidence for large Hydrogen and Helium envelopes containing several percent of the planet's mass and display a large diversity in mean densities. Here we show that this wide range can be achieved by one or two late giant impacts, which are frequently needed to achieve long-term orbital stability in multiple planet systems once the gas disk has disappeared. We demonstrate using hydrodynamical simulations that a single collision between similarly sized exoplanets can easily reduce the envelope-to-core-mass ratio by a factor of two and show that this leads to a corresponding increase in the observed mean density by factors of 2-3. In addition we investigate how envelope-mass-loss depends on envelope mass, planet radius, semi-major axis, and the mass distribution inside the envelope. We propose that a small number of giant impacts may be responsible for the large observed spread in mean densities, especially for multiple-planet systems containing planets with very different densities and which have not been significantly sculpted by photo evaporation.
\end{abstract}


\section{Introduction}
With the number of known exoplanets climbing into the thousands, exoplanet research is at a truly exciting time. It has already been established that the planet occurrence rate per Sun-like star is more than 50\% for planets larger than Earth and smaller than Neptune and with orbital periods of less than about 100 days \citep{F13,PHM13}, which makes these so called super-Earths and mini-Neptunes the most abundant planets in our galaxy known to date. Super-Earths and mini-Neptunes frequently show evidence for large Hydrogen and Helium envelopes containing several percent of the planet's mass \citep{LF13}. Figure 1 displays the mean densities for exoplanets with measured masses and which have radii, $R< 4 R_{\earth}$, where $R_{\earth}$ is the radius of the Earth. Panel a) shows the mean densities as a function of stellar flux, $F$, received by the planets. A large spread in mean densities spanning more than one order of magnitude is apparent for fluxes of less than about 200 $F_{\earth}$, where $F_{\earth}$ is the flux from our Sun at 1~AU. The deficit of low mean densities for $F> 200 F_{\earth}$ is well explained by photo-evaporation which can strip significant fractions of the gaseous envelopes of highly irradiated planets \citep{LF12,LF13B}. And although it has been clearly shown that photo-evaporation can lead to large changes in a planet's mean density and that this may explain the diverse densities in the Kepler 11 system \citep{LF12}, it likely can't account for the large spread in mean densities of planets with low stellar fluxes for which it should not be important. Panel b) in Figure 1 displays the mean densities as a function of planet mass, $M$, where $M_{\earth}$ is the mass of the Earth. It illustrates clearly that super-Earths and mini-Neptunes of a given mass display a large range in mean densities. This is surprising because formation models would naively predict a single mass-radius relationship \citep{IS15,LC15} and one would need to appeal to a diversity in formation environments to account for the large scatter \citep[e.g.][]{D15}. This is especially unsatisfactory for planets in multiple systems which display a large diversity in mean density [e.g. Kepler-20 \citep{F12}, Kepler-36 \citep{C12}, Kepler-48 \citep{MA14,S13}, and Kepler-68 \citep{G13}]. 

In this letter we propose that the large range of observed mean densities may be caused by one or two giant impacts that occurred once the gas disk dissipated. Such giant impacts are expected to be common because they are needed to provide long-term orbital stability of planetary systems and occur typically on timescales between 10 and 100~Myrs \citep{CR14}. We calculate the planetary radii as a function of mass for ages of 10 to 100~Myrs and use these as input parameters in our one-dimensional hydrodynamical simulations in which we calculate the envelope fraction lost due to giant impacts for initial envelope fractions of 1-10\%. We demonstrate  that a single collision between similarly sized exoplanets can easily reduce the envelope-to-core-mass ratio by a factor of two. By following the planets' thermal evolution over several Gyrs, we show that this leads to a corresponding increase in mean densities by factors of 2-3.

This letter is structured as follows. In section 2.1, we construct our own thermal evolution model to calculate the planet radii for a given core mass as a function of time. We show in section 2.2 that giant impacts can significantly reduce the envelope-to-core-mass ratio and demonstrate in section 2.3 that this results in a large increase in a planet's mean density. Our discussions and conclusions follow in section 3.

\citet{LH15} independently proposed the idea that giant impacts may be responsible for the large diversity in exoplanet densities; their paper investigates the mass loss for two specific giant impacts using three-dimensional hydrodynamic simulations and their paper was posted on the arXiv as we were preparing this manuscript for submission. 

\begin{figure}
\centering
\subfigure{\includegraphics[width=100mm]{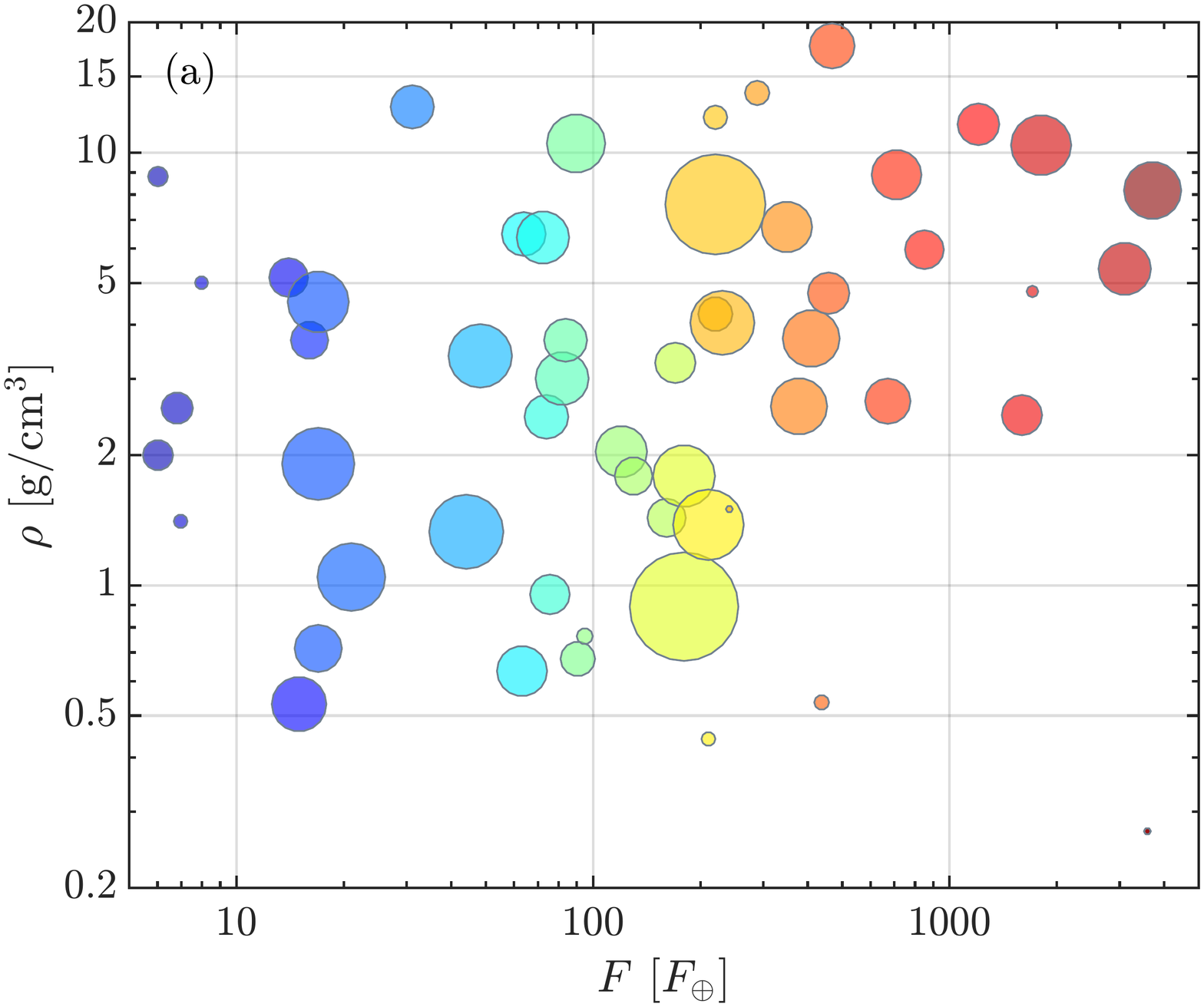}}\\
\subfigure{\includegraphics[width=100mm]{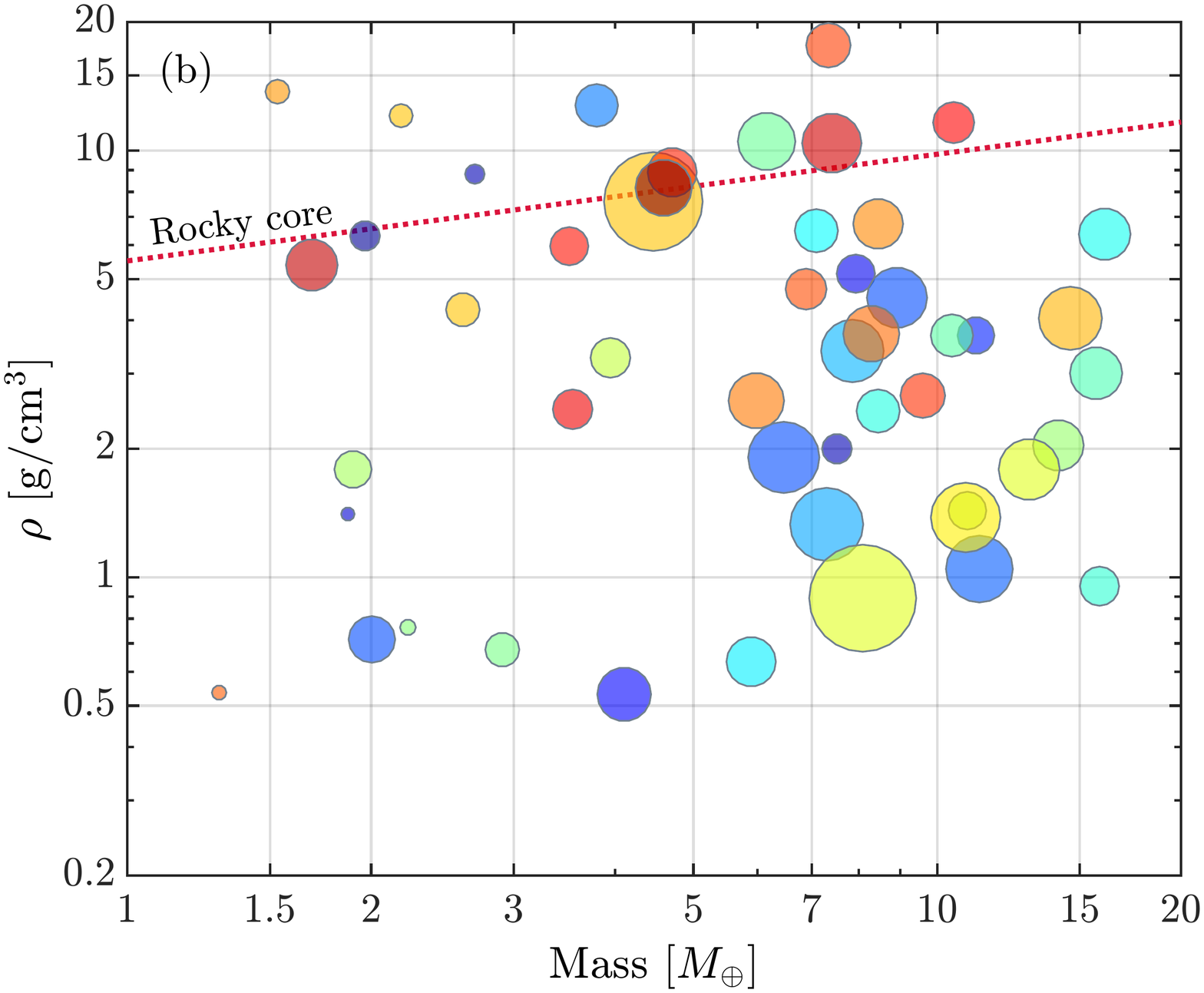}}\\
\subfigure{\includegraphics[width=100mm]{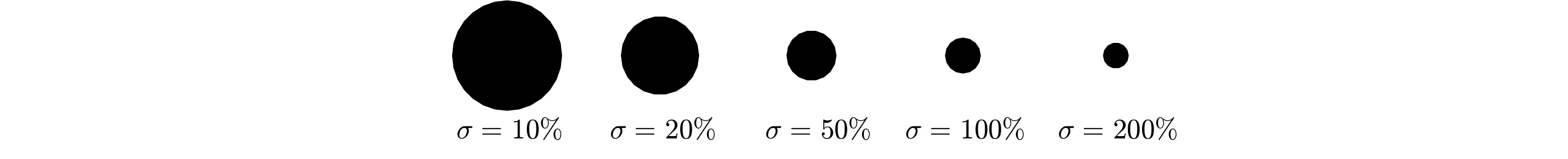}}\\
\caption{Densities of exoplanets with $R<4 R_{\earth}$. The surface area of each data point is inversely proportional to the $1 \sigma$ error of the density estimate, such that the most secure density measurements correspond to the largest points. The normalization of the error bars is shown at the bottom of the figure. The colors of the points represent the amount of flux received from the host star. Panel a) shows mean density as a function of flux, $F$, in units of the Earth flux, $F_{\earth}$. Panel b) displays exoplanet densities as a function of planet mass in units of Earth masses, $M_{\earth}$. Most data are taken from \citet{WM14} and references therein. Additional data taken from \citet{JH15} and \citet{B15}. For reference, a mean density curve assuming a purely rocky planet \citep{SK07} is shown with a dotted red line.}
\label{Fig1}
\end{figure}

\section{Planet Formation \& Late Giant Impacts}
\citet{R14} has shown by modeling the composition of planets with measured radii and masses that the majority of planets larger than $\sim 1.6 R_{\earth}$ have significant gaseous envelopes. This implies that these planets likely formed and interacted with the primordial gas disk. This interaction is expected to have resulted in migration and efficient damping of their eccentricities and inclinations leading to densely packed planetary systems regardless of their exact formation location. As the gas disk disappears on timescales of 1-10~Myrs \citep{H08}, secular excitation in densely packed planetary systems will lead to eccentricity growth culminating in one or two giant impacts producing planetary systems with long term stability \citep{CR14,DPH13}. A large number of multiple planet systems discovered by Kepler may therefore have undergone one or  two large collisions after the gas disk disappeared.

\subsection{Initial Planetary Radii and Thermal Evolution}
The radii of planets with significant gaseous envelopes will shrink with time as the planets cool and their envelopes contract. We evolve the contraction of an envelope of fixed mass $M_{en}$ about a core with mass $M_{c}$ over time using the method outlined in \citet{PY14}. We use the term `core' here to mean the rocky part of the planet. Typically, for such evolutionary models, the entropy of the envelope when contraction begins is set at an initial, high value (``hot start''), and the planet is then allowed to cool  \citep{LF13,HB15}. Here we assume the envelope starts off with an arbitrary high intrinsic luminosity, and we solve the equations of hydrostatic equilibrium iteratively to determine the radius corresponding to a given $(M_{en},M_{c})$ pair. We do the same for a range of luminosities, so that by linking the radius change to a change in internal luminosity and the energy budget of the envelope, we can track the evolution of envelope radius with time. We choose the initial luminosity to be $10^{32}~\mathrm{erg/s}$, although we find the results are generally insensitive to the exact choice of this initial condition for $M_c \gtrsim 2M_{\earth}$ (\citealt{LF13}; see also section 2.3 below). 

\begin{figure}[htp]
\centerline{
\epsfig{file=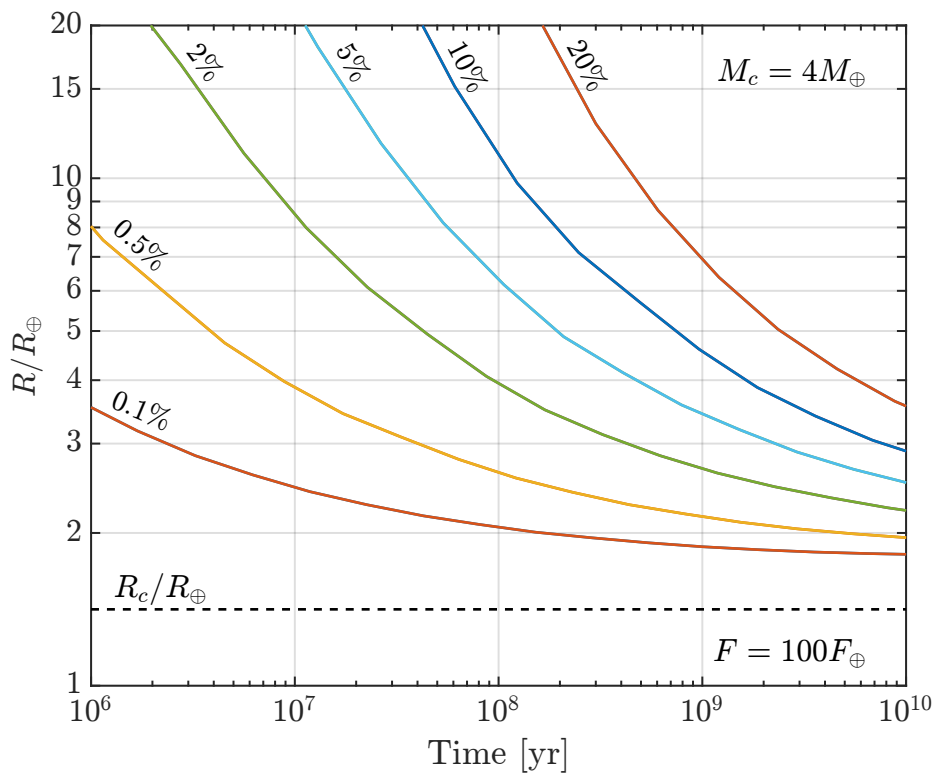, scale=0.5}}
\caption{Radius evolution as a function of time for a planet with $M_c=4M_{\earth}$. The different colored lines correspond to different envelope-to-core-mass ratios and the black-dashed line to the core radius. We assume that the core radius scales with the core mass as $R_c/R_{\earth}=(M_c/M_{\earth})^{1/4}$ \citep{SK07}.}
\label{Fig2}
\end{figure}

For the super-Earths and mini-Neptunes with large Hydrogen-Helium adiabats  the temperature and pressure at the core surface exceed several thousand kelvin and many kilobars, respectively. These values imply that the rocky core should be partially or fully molten enabling easy heat transfer between the rocky core and the gaseous envelope \citep{H09}. Since for super-Earths and mini-Neptunes most of the mass is in there core, the core can contribute significantly to the overall energy budget. For the heat capacity of the core we assume $c_p = 10^{7}~\mathrm{erg~g^{-1}K^{-1}}$ \citep{A02}. We do not include heating from radioactive decay, which we found to have little impact on our results. We set our initial pressure boundary condition to $20~\mathrm{mbar}$, suitable for the viewing geometry of optical transits \citep{LF13}. Our outer temperature boundary condition is set by assuming a stellar flux of $100F_{\oplus}$. The equations of hydrostatic equilibrium are supplemented with those of energy transport. We assume that when the Schwarzschild instability criterion is satisfied, then energy transport is convective, and that when it is not, energy transport is due to radiative diffusion [see \citet{IS15} for further details]. In the latter case, energy transport is governed by the local optical depth. Our opacities are determined from OPAL opacity tables \citep{IR96}. We assume a metallicity of $Z = 0.02$ with Hydrogen and Helium mass fractions of $X = 0.80$ and $Y = 0.18$, respectively, yielding a mean molecular mass of 2.3 proton masses. We ignore mass loss due to photo-evaporation. In Figure \ref{Fig2}, we show an example of the cooling history of a planet with a $4M_{\oplus}$ core and different envelope mass fractions. Envelopes with larger masses contract over longer timescales since they have a larger energy budget and hence longer Kelvin-Helmholtz timescales. 

\subsection{Envelope-Mass-Loss due to a Giant Impact}
Using one dimensional hydrodynamical simulations, we calculated the envelope-mass-loss resulting from a giant impact. We only model the adiabatic part of the envelope, since the thin isothermal outer-layer contains negligible mass. We track the propagation of a shock launched into the envelope due to local ground motion by solving the hydrodynamic equations with a finite-difference, Lagrangian scheme. Each mass parcel is tracked, and if it reaches velocities greater than its initial, radius-dependent escape velocity from the planet, it is considered lost. We determine the global envelope-mass-loss fraction by integrating the local mass loss over the entire surface of the planet where we account for the global distribution of the different ground velocities. The ground velocities were calculated by relating the impactor mass, $m,$ and impact velocity, $v_{imp}$, to
the resulting ground motion at the various locations of the planet by approximating the impacts as point like explosions on a sphere. Such explosions result in self-similar solution of the second type \citep{ ZR67}. Assuming momentum conservation \citep{LS12} we find that the velocity component of the shocked fluid perpendicular to the planet's surface is given by
\begin{equation}
v_g=v_{imp} \left(\frac{m}{M}\right)\frac{1}{(l/2R)^2[4-3(l/2R)]},
\end{equation}
\citep{SS15} where $R$ is the radius of the planet and $l$ the distance that the shock has transversed from the impact point. We compared our envelope-mass-loss results with those reported in \citet{SL14} and \citet{LS14} who used 3D impact simulations to determine the surface velocity field. For the parameters corresponding to the various Moon-forming scenarios investigated in their work, we find good agreement between their envelope-mass-loss results and ours. Further details of our model can be found in \citet{SS15}. 

Since the impacts are triggered once the gas disk has disappeared, they typically happen when the planetary system is between 10-100~Myrs old. As shown in Figure \ref{Fig2}, the planetary radii at these early times are significantly more extended than at ages of a few Gyrs by when they had time to cool and contract \citep[e.g.][]{LF13}. When estimating the envelope-mass-loss fraction due to a planetary collision we therefore determine the loss fraction for a range of planetary radii. We note here that a giant impact can significantly modify the radial profile of the envelope. However, since the envelope profile after thermal evolution over Gyr timescales is generally insensitive to the exact conditions during the first few tens of millions of years, the collision is not expected to leave any significant long-term signatures in the planet's envelope. 

\begin{figure}[htp]
\centerline{
\epsfig{file=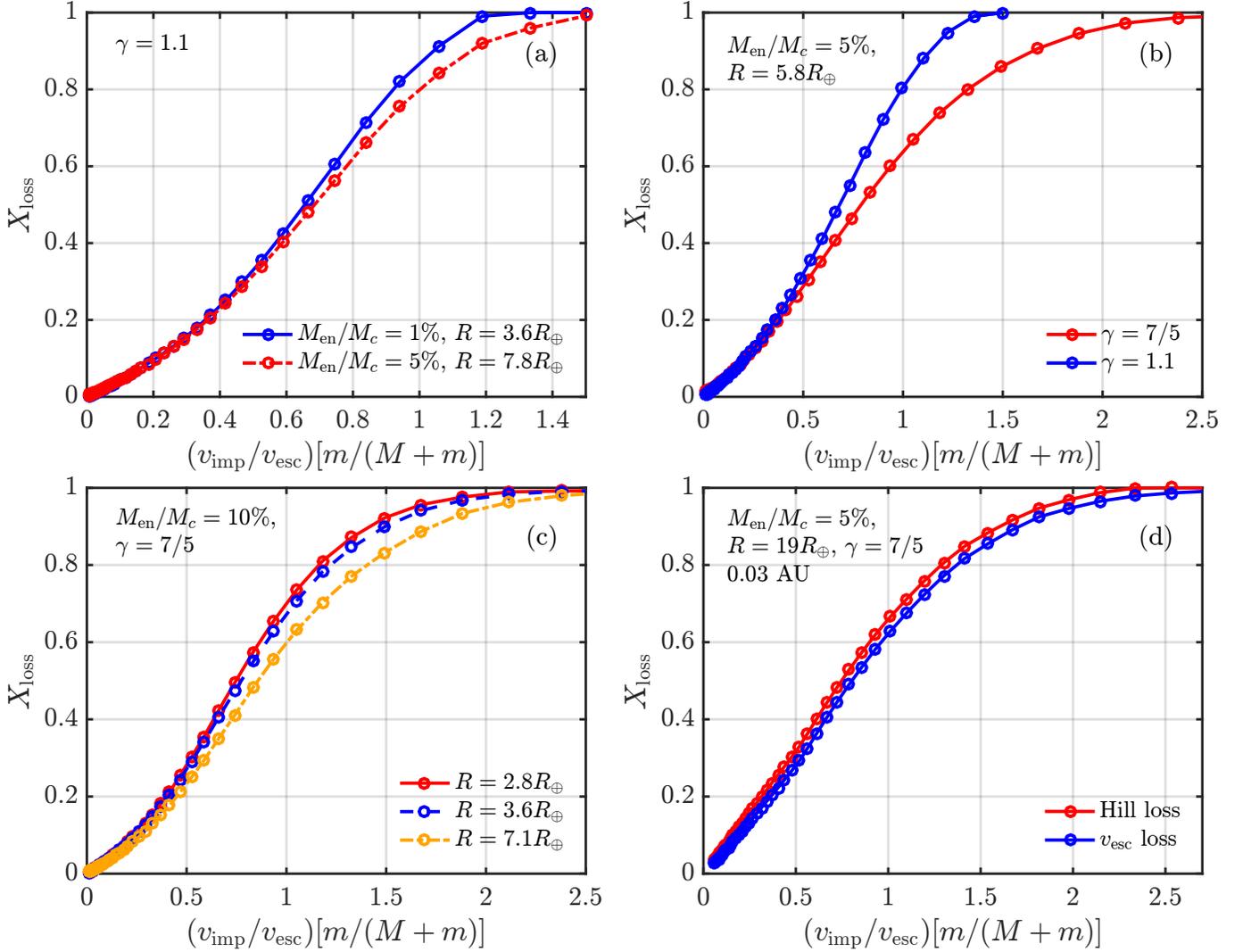, scale=0.7}}
\caption{Global envelope-mass-loss fraction, $X_{loss}$, as a function of impact momentum, $(v_{imp}/v_{esc})[m/(M+m)]$, where $v_{imp}$ and $m$ are the impact velocity and mass of the impactor, and $v_{esc}$ and $M$ are the mutual escape velocity and mass of the target, respectively. The global mass loss fraction was obtained by summing the envelope fraction lost over the whole surface of the core. Panel a) gives the envelope fraction lost for envelope mass fractions of 1\% and 5\%, where the planetary radii were chosen such that they correspond to systems that are 50~Myrs of age; panel b) shows how the adiabatic index, $\gamma$, which determines the mass distribution inside the envelope, affects the results; panel c) displays how the mass loss depends on the radius of the planet; panel d) demonstrates increased atmospheric loss for planets that are very close to their host star due to their small Hill-radii (see section 2.2 for details).
A $4M_{\earth}$ core was assumed in all cases, but we found that the envelope-mass-loss fraction depends only very weakly on core mass.}
\label{Fig3}
\end{figure}

Figure \ref{Fig3}a shows the resulting mass loss as a function of impactor momentum from our hydrodynamical simulations for envelope mass fractions of 1\% and 5\%. The planetary radii were chosen such that they correspond to systems that are 50~Myrs of age (see Figure \ref{Fig2}). An adiabatic index of $\gamma=1.1$ was used in this simulation because when examining the planet's thermal and density profiles during the accretion and cooling phase, it has been found that $\gamma<4/3$ \citep{LC15}. This low value of $\gamma$ is due to the dissociation of hydrogen, which we find is marginally important for super-Earths and mini-Neptunes. The value of $\gamma$ is interesting because it determines how the mass is distributed inside the gaseous envelope. 
Since the density, $\rho$, of an envelope dominated by convection scales as $\rho \propto z^{-1/(\gamma-1)}$, we find that its mass is given by
\begin{equation}
M_{en} \propto z^{\frac{3\gamma-4}{\gamma-1}},
\end{equation} 
where $z$ is the height in the envelope. Hence, for $\gamma<4/3$, the mass in an adiabatic envelope is concentrated towards the core, whereas for values of $\gamma>4/3$, which applies for diatomic gas with 5 degrees of freedom ($\gamma=7/5$) and monoatomic gas with three degrees of freedom ($\gamma=5/3$), the mass is concentrated towards the radiative-convective boundary. How the mass distribution in the gaseous envelope affects the global mass loss is shown in Figure \ref{Fig3}b, which displays the envelope-mass-loss fraction for $\gamma=7/5$ and $\gamma=1.1$. For identical planet masses and radii more mass is lost for the $\gamma=1.1$ than the $\gamma=7/5$ case because in the former the mass of the envelope is concentrated towards the core such that the shock that is launched into the envelope from the core can impart a larger momentum onto the envelope. Figure \ref{Fig3}c displays the mass loss dependence on the envelope radius. For $\gamma=7/5$ we find that for identical collision parameters less mass is lost for larger envelope radii. This result arises because for $\gamma=7/5$ the envelope mass is concentrated towards the edge of the envelope and larger radii result in lower envelope densities at the core, which in turn implies that the shock travels with a smaller momentum into the envelope. In contrast, for the $\gamma = 1.1$ case, the mass loss dependence on envelope radius would be weaker, because most of the envelope mass is concentrated towards the core. Finally, Figure \ref{Fig3}d shows how much the envelope-mass-loss could be increased for very close-in planets because of their small Hill radii. When calculating the envelope-mass-loss fraction a fluid parcel is lost when it was accelerated to velocities greater or equal to its original escape velocity. For very close-in planets the mass loss can be increased because it is sufficient for a fluid parcel to reach velocities to reach to the Hill radius, $R_H=a (M/3M_{\star})^{1/3}$, where $a$ is the semi-major axis and $M_{\star}$ the mass of the star. The velocity needed for escape is given by
\begin{equation}
v=v_{esc}(z)\sqrt{1-z/R_H}
\end{equation}
where $v_{esc}(z)$ is the escape velocity at a given position $z$ in the envelope before the impact and $z$ is measured from the center of the core. Usually, the Hill radius is much larger than the radius of the planet such that $v=v_{esc}(z)$, but for extended envelopes and small semi-major axis the planets radius can become comparable to the Hill radius. An example of such a case is shown in Figure \ref{Fig3}d, where the planet's radius is chosen such that it is equal to its Hill radius.

Using the results presented in Figure \ref{Fig3} we can read off the envelope-mass-loss fraction, $X_{loss}$, for given impact parameters. For an impactor with mass $m$ and radius $r$ and target of mass $M$ and radius $R$ the impact velocity is given by  $v_{imp}=\sqrt{v_{\infty}^2 + v_{esc}^2}$, where  $v_{esc} \equiv \sqrt{2G(M+m)/(R+r)}$ is the mutual escape velocity. We therefore find that a collision between comparable mass planets with $v_{\infty} \sim v_{esc}$ that about half of the gaseous envelope of both target and impactor is lost. This yields a final planet with a core mass that is about twice the original mass and an envelope-to-core-mass ratio that is reduced by a factor of two.

\subsection{Relating Envelope Mass Fractions to Mean Densities}
To relate this reduction in $M_{en}/M_{c}$ to observed mean densities we obtained the mean density for planets of various masses for different $M_{en}/M_{c}$. We did this by calculating the contraction of the planet's radius as a function of time and then obtained for a given planet mass and $M_{en}/M_{c}$ the corresponding radius at an age of 1 Gyr (see section 2.1 details). 

\begin{figure}
\centering
\subfigure{\includegraphics[width=110mm]{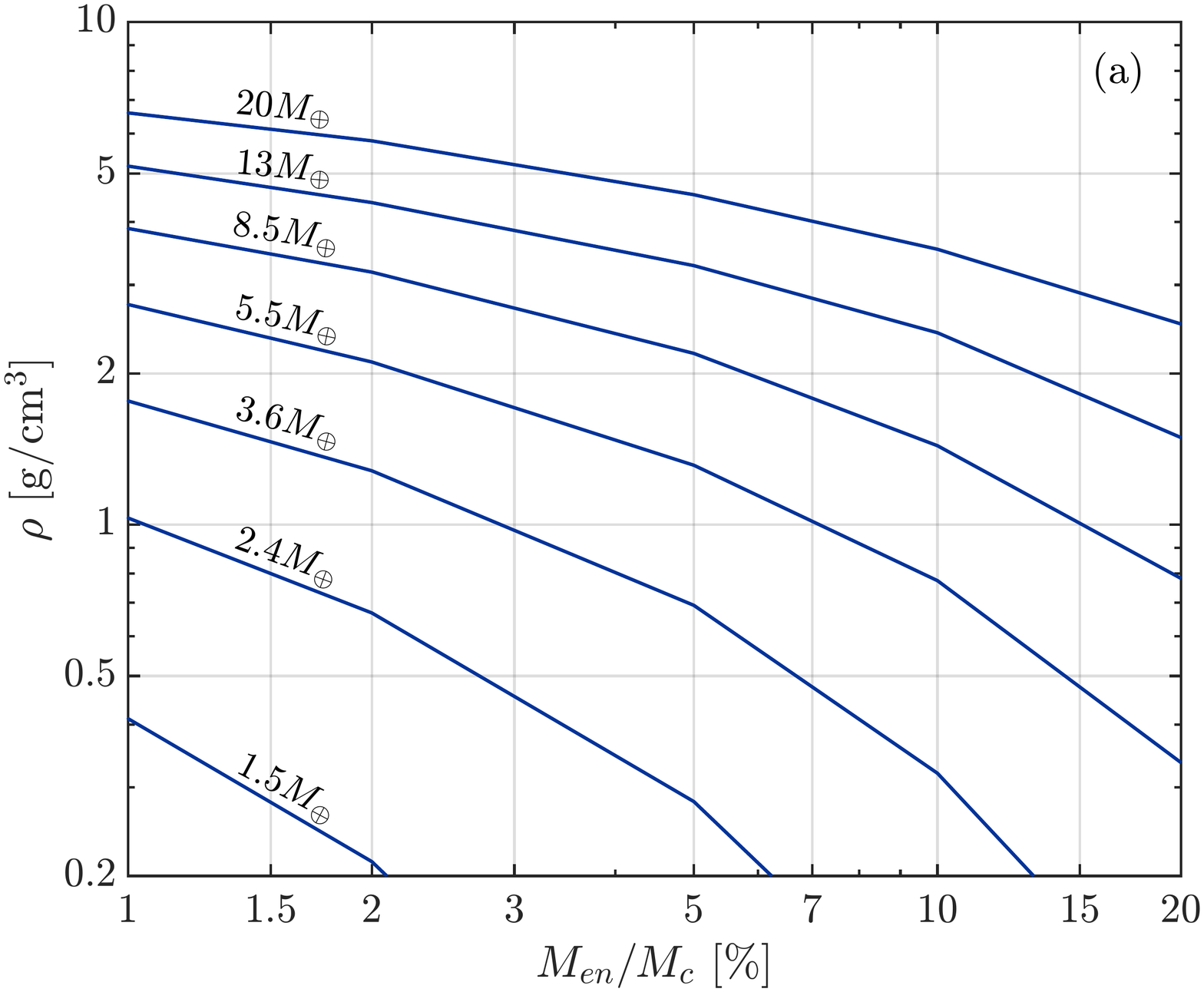}}\\
\subfigure{\includegraphics[width=110mm]{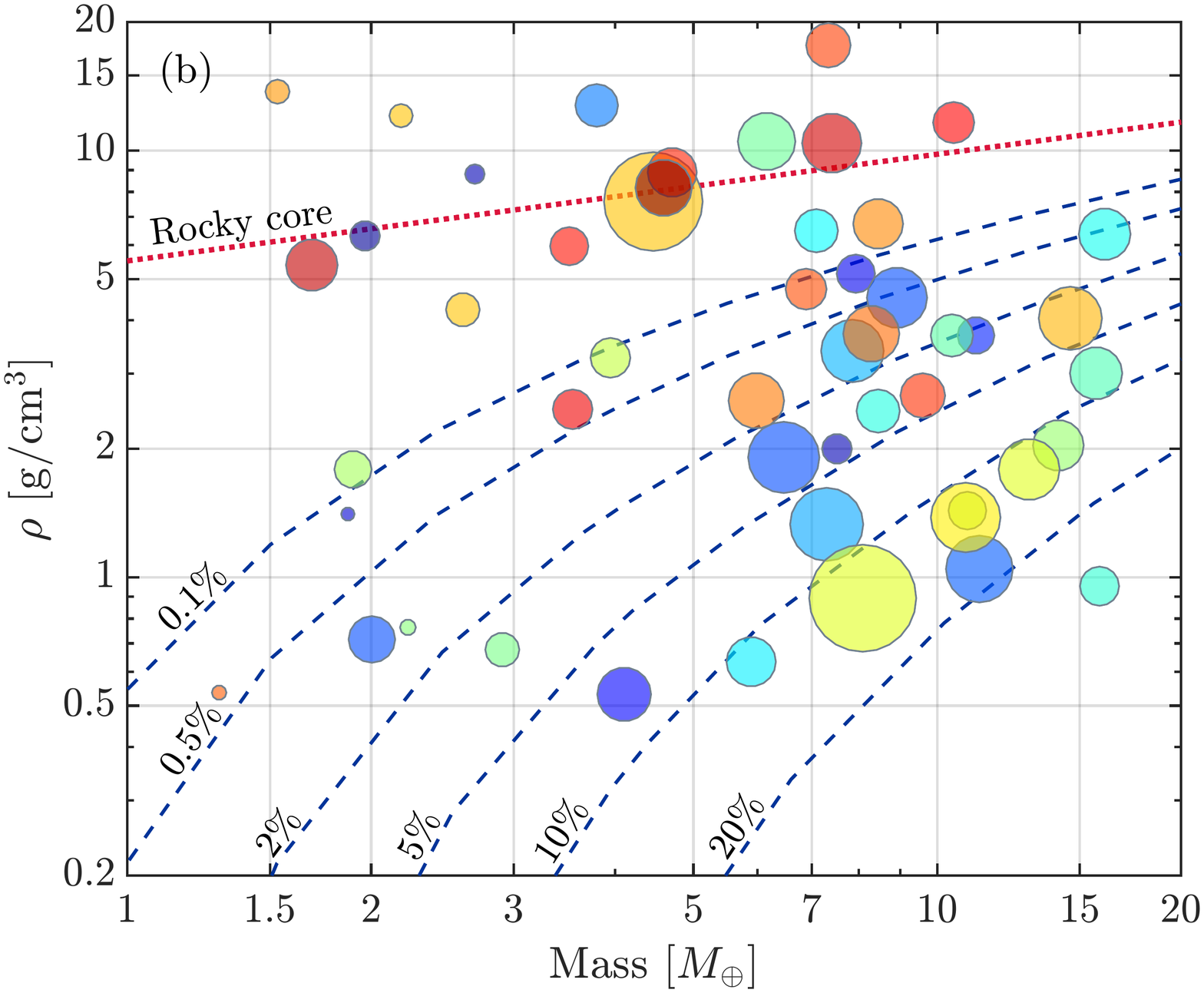}}\\
\caption{Exoplanet mean densities, $\rho$, as a function of mass and envelope-to-core-mass fraction, $M_{en}/M_c$. Panel a) displays the density as a function of $M_{en}/M_c$; the dashed lines correspond to different planet masses which are given adjacent to each line. The dashed lines were calculated numerically by following the thermal contraction of the planet's envelope over 1 Gyr for a flux of $100F_{\earth}$. Panel b) shows the mean density as a function of planet mass. Each blue line represents the envelope-to-core-mass fraction of its label, the dotted red line corresponds to a rocky core without any atmosphere, and the points correspond to the same exoplanet data displayed in Figure \ref{Fig1}.}
\label{Fig4}
\end{figure}

Figure \ref{Fig4}a displays the resulting exoplanet densities as a function of $M_{en}/M_{c}$ and planet mass from our thermal evolution models.  It shows that a reduction of $M_{en}/M_{c}$ by a factor of 2 due to a comparable mass merger results in a corresponding increase in the mean density by factors of 2-3. For instance, a planet with $M_{c} = 4M_{\earth}$ and $M_{en}/M_{c} = 5\%$ has a mean density of 0.8 $\mathrm{g/cm^3}$.  After a giant impact with an equal mass core that ejects half the envelope mass (such that $M_{c} = 8M_{\earth}$ and $M_{en}/M_{c} = 2.5\%$), the mean density is $\rho = 2.8~\mathrm{g/cm^3}$, which is an increase by a factor of 3.5. For a similar collision but for an initial atmosphere of $2\%$, the resulting mean density of the final planet is increased by a factor of 2. This demonstrates that one or two giant impacts can give rise to a large spread in mean densities. In Figure \ref{Fig4}b, we show a comparison of planet mass vs. mean density of observed exoplanet systems and display as blue dashed lines density contours for various envelope-to-core-mass ratios. The mean densities we find after $1~\mathrm{Gyr}$ of cooling agree with those of \citet{LF13} typically to within about 10-25\% for core masses $\gtrsim 2M_{\oplus}$. At lower core masses $\lesssim 2M_{\oplus}$, we find, similar to \citet{HB15}, somewhat larger planetary radii than reported by \citet{LF13}. Specifically, we find radii that are up to 50\% larger than those calculated by \citet{LF13} and which result in lower mean densities for small planets. We suspect that these discrepancies are likely due to the different initial conditions used [fixed luminosity in this work, fixed entropy in \citet{LF13}, and fixed radii in \citet{HB15}] all of which have the greatest impact on the thermal evolution of low mass planets $\lesssim 2M_{\earth}$. We note that our density results for Neptune-mass planets with low mean densities suggest that these planets possess roughly 20\% of their mass in gaseous envelopes. This is interesting because planets with envelope mass fractions in excess of about 20\% are expected to undergo runaway gas accretion leading to the formation of a gas giant instead \citep{R06,PY14}.  

\section{Discussion and Conclusions}
We have shown here that giant impacts between similarly-sized planets can easily reduce the envelope-to-core-mass ratio by factors of two and that this leads to an increase in mean density by factors of 2-3. Since late giant impacts are frequently needed to achieve long-term orbital stability in multiple planet systems once the gas disk has disappeared \citep{CR14}, we propose here that a small number of giant impacts may have given rise to the large spread observed in mean densities. Furthermore, giant impacts naturally yield a large diversity in densities in a given planetary system, because they are stochastic in nature and because typically only a small number of giant impacts is needed to achieve long-term stability. We suggest that the observed diversity in densities amongst members of multiplanet systems such as Kepler-11, Kepler-20, Kepler-36, Kepler-48, and Kepler-68 may be the result of such late-stage giant impacts.

The envelope-mass-loss that we calculate in section 2 may be enhanced by photo-evaporation and via a Parker wind \citep{LH15} both of which should most strongly affect planets with low surface gravity on small semi-major axis \citep{LF13B}. As the number of mass measurements increases for planets at larger semi-major axis one should be able to disentangle the contribution of envelope-mass-loss due to or enhanced by photo-evaporation and that due to collisions. The fact that Figure 1 shows a large spread in mean densities for low stellar fluxes already suggests that photo-evaporation is likely not the main cause for this diversity.

\acknowledgements
We thank the anonymous referee for constructive comments that helped improve the manuscript. HS thanks Re'em Sari and Sivan Ginzburg for helpful discussions.

\bibliographystyle{aj} 

\begin{thebibliography}{26}
\expandafter\ifx\csname natexlab\endcsname\relax\def\natexlab#1{#1}\fi
\expandafter\ifx\csname url\endcsname\relax
  \def\url#1{\texttt{#1}}\fi
\expandafter\ifx\csname urlprefix\endcsname\relax\def\urlprefix{URL }\fi
\providecommand{\eprint}[2][]{\url{#2}}

\bibitem[{{Alf{\`e}} et~al.(2002){Alf{\`e}}, {Price}, \& {Gillan}}]{A02}
{Alf{\`e}}, D., {Price}, G.~D., \& {Gillan}, M.~J. 2002, \prb, 65, 165118.
  \eprint{cond-mat/0107307}
  
\bibitem[{{Barros}, {Almenara}, {Demangeon} et~al.(2015)}]{B15}
{Barros}, S.~C.~C., {Almenara}, J.~M., {Demangeon}, O., et~al. 2015, \mnras,
  454, 4267. \eprint{1510.01047}

\bibitem[{{Carter}, {Agol}, {Chaplin} et~al.(2012)}]{C12}
{Carter}, J.~A., {Agol}, E., {Chaplin}, W.~J., et~al. 2012, Science, 337, 556.
  \eprint{1206.4718}  
 
\bibitem[{{Cossou} et~al.(2014){Cossou}, {Raymond}, {Hersant}, \&
  {Pierens}}]{CR14}
{Cossou}, C., {Raymond}, S.~N., {Hersant}, F., \& {Pierens}, A. 2014, \aap,
  569, A56. \eprint{1407.6011}

\bibitem[{{Dawson} et~al.(2015){Dawson}, {Chiang}, \& {Lee}}]{D15}
{Dawson}, R.~I., {Chiang}, E., \& {Lee}, E.~J. 2015, \mnras, 453, 1471.
  \eprint{1506.06867}

\bibitem[{{Deck} et~al.(2013){Deck}, {Payne}, \& {Holman}}]{DPH13}
{Deck}, K.~M., {Payne}, M., \& {Holman}, M.~J. 2013, \apj, 774, 129.
  \eprint{1307.8119}

\bibitem[{{Fressin}, {Torres}, {Rowe} et~al.(2012)}]{F12}
{Fressin}, F., {Torres}, G., {Rowe}, J.~F., et~al. 2012, \nat, 482, 195.
  \eprint{1112.4550}  

\bibitem[{{Fressin} et~al.(2013){Fressin}, {Torres}, {Charbonneau}, {Bryson},
  {Christiansen}, {Dressing}, {Jenkins}, {Walkowicz}, \& {Batalha}}]{F13}
  {Fressin}, F., {Torres}, G., {Charbonneau}, D., {Bryson}, S.~T.,
  {Christiansen}, J., {Dressing}, C.~D., {Jenkins}, J.~M., {Walkowicz}, L.~M.,
  \& {Batalha}, N.~M. 2013, \apj, 766, 81. \eprint{1301.0842}

\bibitem[{{Gilliland}, {Marcy}, {Rowe} et~al.(2013)}]{G13}
{Gilliland}, R.~L., {Marcy}, G.~W., {Rowe}, J.~F., et~al. 2013, \apj, 766, 40.
  \eprint{1302.2596}  

\bibitem[{{Henning} et~al.(2009){Henning}, {O'Connell}, \& {Sasselov}}]{H09}
{Henning}, W.~G., {O'Connell}, R.~J., \& {Sasselov}, D.~D. 2009, \apj, 707,
  1000. \eprint{0912.1907}

\bibitem[{{Hillenbrand}(2008)}]{H08}
{Hillenbrand}, L.~A. 2008, Physica Scripta Volume T, 130, 014024.
  \eprint{0805.0386}

\bibitem[{{Howe} \& {Burrows}(2015)}]{HB15}
{Howe}, A.~R., \& {Burrows}, A. 2015, \apj, 808, 150. \eprint{1505.02784}

\bibitem[{{Iglesias} \& {Rogers}(1996)}]{IR96}
{Iglesias}, C.~A., \& {Rogers}, F.~J. 1996, \apj, 464, 943

\bibitem[{{Inamdar} \& {Schlichting}(2015)}]{IS15}
{Inamdar}, N.~K., \& {Schlichting}, H.~E. 2015, \mnras, 448, 1751.
  \eprint{1412.4440}

\bibitem[{{Jontof-Hutter} et~al.(2015){Jontof-Hutter}, {Rowe}, {Lissauer}, {Fabrycky} \& {Ford}}]{JH15}
{Jontof-Hutter}, D.~G., {Rowe}, J.~F., {Lissauer}, J.~J., {Fabrycky}, D.~C., \& {Ford}, E.~B. 2015, \nat, 522,
  321. \eprint{1506.07067}
  
\bibitem[{{Lee} \& {Chiang}(2015)}]{LC15}
{Lee}, E.~J., \& {Chiang}, E. 2015, \apj, 811, 41. \eprint{1508.05096}

\bibitem[{{Leinhardt} \& {Stewart}(2012)}]{LS12}
{Leinhardt}, Z.~M., \& {Stewart}, S.~T. 2012, \apj, 745, 79. \eprint{1106.6084}

\bibitem[{{Liu} et~al.(2015){Liu}, {Hori}, {Lin}, \& {Asphaug}}]{LH15}
{Liu}, S.-F., {Hori}, Y., {Lin}, D.~N.~C., \& {Asphaug}, E. 2015, ArXiv
  e-prints. \eprint{1509.05772}

\bibitem[{{Lock} et~al.(2014){Lock}, {Stewart}, \& {Mukhopadhyay}}]{LS14}
{Lock}, S.~J., {Stewart}, S.~T., \& {Mukhopadhyay}, S. 2014, in Lunar and
  Planetary Science Conference, vol.~45 of Lunar and Planetary Science
  Conference, 2843

\bibitem[{{Lopez} \& {Fortney}(2013)}]{LF13B}
{Lopez}, E.~D., \& {Fortney}, J.~J. 2013, \apj, 776, 2. \eprint{1305.0269}

\bibitem[{{Lopez} \& {Fortney}(2014)}]{LF13}
--- 2014, \apj, 792, 1. \eprint{1311.0329}

\bibitem[{{Lopez} et~al.(2012){Lopez}, {Fortney}, \& {Miller}}]{LF12}
{Lopez}, E.~D., {Fortney}, J.~J., \& {Miller}, N. 2012, \apj, 761, 59.
  \eprint{1205.0010}

\bibitem[{{Marcy}, {Isaacson}, {Howard} et~al.(2014)}]{MA14}
{Marcy}, G.~W., {Isaacson}, H., {Howard}, A.~W., et~al. 2014, \apjs, 210, 20.
  \eprint{1401.4195}  

\bibitem[{{Petigura} et~al.(2013){Petigura}, {Howard}, \& {Marcy}}]{PHM13}
{Petigura}, E.~A., {Howard}, A.~W., \& {Marcy}, G.~W. 2013, Proceedings of the
  National Academy of Science, 110, 19273. \eprint{1311.6806}

\bibitem[{{Piso} \& {Youdin}(2014)}]{PY14}
{Piso}, A.-M.~A., \& {Youdin}, A.~N. 2014, \apj, 786, 21. \eprint{1311.0011}

\bibitem[{{Rafikov}(2006)}]{R06}
{Rafikov}, R.~R. 2006, \apj, 648, 666

\bibitem[{{Rogers}(2014)}]{R14}
{Rogers}, L.~A. 2014, ArXiv e-prints. \eprint{1407.4457}
  
\bibitem[{{Schlichting} et~al.(2015){Schlichting}, {Sari}, \&
  {Yalinewich}}]{SS15}
{Schlichting}, H.~E., {Sari}, R., \& {Yalinewich}, A. 2015, \icarus, 247, 81.
  \eprint{1406.6435}

\bibitem[{{Seager} et~al.(2007){Seager}, {Kuchner}, {Hier-Majumder}, \&
  {Militzer}}]{SK07}
{Seager}, S., {Kuchner}, M., {Hier-Majumder}, C.~A., \& {Militzer}, B. 2007,
  \apj, 669, 1279. \eprint{0707.2895}

\bibitem[{{Steffen}, {Fabrycky}, {Agol} et~al.(2013)}]{S13}
{Steffen}, J.~H., {Fabrycky}, D.~C., {Agol}, E., et~al. 2013, \mnras, 428,
  1077. \eprint{1208.3499}  

\bibitem[{{Stewart} et~al.(2014){Stewart}, {Lock}, \& {Mukhopadhyay}}]{SL14}
{Stewart}, S.~T., {Lock}, S.~J., \& {Mukhopadhyay}, S. 2014, in Lunar and
  Planetary Science Conference, vol.~45 of Lunar and Planetary Science
  Conference, 2869

\bibitem[{{Weiss} \& {Marcy}(2014)}]{WM14}
{Weiss}, L.~M., \& {Marcy}, G.~W. 2014, \apjl, 783, L6. \eprint{1312.0936}

\bibitem[{{Zel'dovich} \& {Raizer}(1967)}]{ZR67}
{Zel'dovich}, Y.~B., \& {Raizer}, Y.~P. 1967, {Physics of shock waves and
  high-temperature hydrodynamic phenomena}

\end{thebibliography}

\end{document}